\documentstyle[aps,prl,multicol,epsf]{revtex}
\def \BE {\begin{equation}}
\def \EE {\end{equation}}
\def \BEA {\begin{eqnarray}}
\def \EEA {\end{eqnarray}}
\def \CR {\nonumber \\}

\begin{document}

\title{``Noisy'' spectra, long correlations and intermittency in wave turbulence}
\author{Yuri V. Lvov$^\dagger$ and Sergey Nazarenko$^*$ 
}
\address
{ 
$^\dagger$ 
Department of Mathematical Sciences, Rensselaer Polytechnic Institute,
Troy, NY 12180 \\ 
$^*$
Mathematics Institute, The University of Warwick,  Coventry, CV4-7AL, UK}

\maketitle







\begin{abstract}
We study the k-space fluctuations of the waveaction about its mean
spectrum in the turbulence of dispersive waves. We use a minimal model
based on the Random Phase Approximation (RPA) and derive evolution
equations for the arbitrary-order one-point moments of the wave
intensity in the wavenumber space. The first equation in this series
is the familiar Kinetic Equation for the mean waveaction spectrum,
whereas the second and higher equations describe the fluctuations
about this mean spectrum. The fluctuations exhibit a nontrivial
dynamics if some long coordinate-space correlations are present in the
system, as it is the case in typical numerical and laboratory
experiments. Without such long-range correlations, the fluctuations
are trivially fixed at their Gaussian values and cannot evolve even if
the wavefield itself is non-Gaussian in the coordinate space.  Unlike
the previous approaches based on smooth initial k-space cumulants, the
RPA model works even for extreme cases where the k-space fluctuations
are absent or very large and intermittent.  We show that any initial
non-Gaussianity at small amplitudes propagates without change toward
the high amplitudes at each fixed wavenumber.  At each fixed
amplitude, however, the PDF becomes Gaussian at large time.

\end{abstract}


\begin{multicols}{2}

\section{Introduction}
The concept of Wave Turbulence (WT), which describes an ensemble of
weakly interacting dispersive waves, significantly enhanced our
understanding of the spectral energy transfer in complex systems like
the ocean, the atmosphere, or in
plasmas~\cite{ZLF,Ben,GS,Zakfil,hasselman}.  This theory also became a
subject of renewed interest recently, (see,
e.g. \cite{OsbornePRL,ZakharovPRL,SoomerePRL,LT}).  Traditionally, WT
theory deals with derivation and solutions of the Kinetic Equation
(KE) for the mean waveaction spectrum (see e.g. \cite{ZLF}).  However,
all experimentally or numerically obtained spectra are ``noisy'', i.e.
exhibit k-space fluctuations which contain a complimentary to the mean
spectra information.  These k-space fluctuations always develop in
numerical experiments even though, typically, most numerical
experiments (e.g. \cite{OsbornePRL,ZakharovPRL}) start with initial
waveaction fields in the k-space which have random phases but which
have no amplitude fluctuations. How fast and why do these amplitude
fluctuations get developed? Are they a numerical artifact or a real
physical phenomenon? Are they Gaussian or some intermittent bursts of
Fourier amplitudes can be expected? These questions remain unanswered
because the waveaction fluctuation have not been studied before.  Such
a study involves description of the higher one-point moments of the
Fourier amplitudes and it will be the main focus of the present
work. We will show that when these one-point moments are not Gaussian
the coordinate space fields are long-correlated. Such fields are very
common in WT, and the numerical initial conditions discussed above is
a typical example. Thus, we will have to generalize WT to include such
long correlated fields.

\section{Random phases vs Gaussian fields. }
The random phase approximation (RPA) has been popular in WT because it
allows a quick derivation of KE \cite{ZLF,GS}.  We will use RPA in
this paper because it provides a minimal model for for description of
the k-space fluctuations of the waveaction about its mean spectrum,
but we will also discuss relation to the approach of \cite{Ben} which
does not assume RPA.

Let us consider a wavefield $a({\bf x}, t)$ in a periodic
box\footnote{ Periodic box is an essential intermediate step for
formulating RPA and for defining the new correlators $M^{(p)}_k$ later
in this paper.  This is related to the fact that, strictly speaking,
the infinite-space Fourier transform is a distribution, rather than a
smooth function, for the class of functions corresponding to
statistically homogeneous fields. The previous theory considered a
class of correlators which are box-size independent and which could be
formally obtained via a direct manipulation with the infinite-box
$a_k$'s.}  of volume ${\cal V}$ and let the Fourier transform of this
field be $a_k$.  Later, we take the large box limit in order to
consider homogeneous wave turbulence. Let us write the complex
function $a_k$ as $a_k =A_k \psi_k $ where $A_k$ is a real positive
amplitude and $\psi_k $ is a phase factor which takes values on the
unit circle centered at zero in the complex plane.  By definition, RPA
for an ensemble of complex fields $a_k$ means that
\begin{enumerate}
\item The phase factors $\psi_k$ are
uniformly distributed on the unit circle in the complex plane
and are statistically independent
of each other 
$$
\langle \psi_{k_1} \psi_{k_2} \rangle =  \delta^1_2, \;\; 
$$
where $\delta^1_2$ is the Kronecker symbol.
\item The phases are statistically independent from the amplitudes $A_k$, 
$$\langle \psi_{k_1} A_{k_2} \rangle = 0$$ Thus, the averaging over
the phase and over the amplitude statistics can be performed
independently.
\item the fluctuations of the amplitudes $A_k$ must also be
decorrelated at different $k$'s.
$$\langle
A_{k_1}^n A_{k_2}^m \rangle =\langle A_{k_1}^n \rangle \langle
A_{k_2}^m \rangle \;\; (m,n = 1,2,3,...).$$
Properties 2 and 3 have typically not been mentioned explicitly before. The
name RPA itself does not refer to the amplitudes but to the phases
only.  However, this important assumption about the amplitude
statistics has always been made implicitely when using RPA, often
without even realizing it.
\end{enumerate}

To illustrate the relation between the random phases and Gaussianity,
let us consider the fourth-order moment for which RPA gives
\begin{equation}
\langle a_{k_1} a_{k_2} {\overline a_{k_3}} {\overline a_{k_4}}
\rangle = n_{k_1} n_{k_2} (\delta^{1}_{3} \delta^{2}_{4} + 
\delta^{1}_{4}\delta^2_3) + Q_{k_1} \delta^1_{2}
\delta^{1}_{3} \delta^{1}_{4},
\label{wick}
\end{equation}
where 
$$ n_k = \langle A_{k}^2 \rangle $$
is the waveaction spectrum and
$$ Q_k = \langle A_{k}^4 \rangle - 2 \langle A_{k}^2 \rangle $$
is a cumulant coefficient.  The last term in this expression appears
because the phases drop out for ${\bf k_1}={\bf k_2}={\bf k_3}={\bf
k_4} $ and their statistics poses no restriction on the value of this
correlator at this point. This cumulant part of the correlator can be
arbitrary for a general random-phased field whereas for Gaussian
fields $ Q_k$ must be zero. Such a difference between the Gaussian and
the random-phased fields occurs only at a vanishingly small set of
modes with ${\bf k_1}={\bf k_2}={\bf k_3}={\bf k_4} $ and it has been typically ignored before
because its contribution to KE is negligible.  Therefore, if the mean
waveaction spectrum was the only thing we were interested in, we could
safely ignore contributions from all (one-point) moments
$$M^{(p)}_k = \langle |a_{k}|^{2p} \rangle \;\; (p=1,2,3,..).$$

However, it is precisely moments $M^{(p)}_k$ that contain information
about fluctuations of the waveaction about its mean spectrum. For
example, the standard deviation of the waveaction from its mean is
\BE
\xi_k = (\langle |a_{k}|^4 \rangle - \langle |a_{k}|^2
\rangle^2)^{1/2} = (M^{(2)}_k - n_k^2)^{1/2}
\label{NazarenkoXI}
\EE
This quantity can be arbitrary for a general random-phased field
whereas for a Gaussian wave field the fluctuation level $\xi_k$ is
fixed, $\xi_k = n_k$.  Note that different values of moments
$M^{(p)}_k$ can correspond to hugely different typical wave field
realizations.  In particular, if $M^{(p)} = n^p$ then there is no
fluctuations and $A_k$ is deterministic, $\xi_k=0$.  For the
opposite extreme of large fluctuations we would have $M^{(p)} \gg n^p$
which means that the typical realization is sparse in the k-space and
is characterized by few intermittent peaks of $A_k$ and close to zero
values in between these peaks. Such an information about the
spectral fluctuations of the waveaction contained in the one-point
moments $M^{(p)}$ is completely erased from the multiple-point moments
by the random phases and it is precisely why these new objects play a
crucial role for the description of the fluctuations.

Will the waveaction fluctuations appear if they were absent initially?
Will they saturate at the Gaussian level $\xi_k = n_k$ or will they
keep growing leading to the k-space intermittency? To answer these
questions, we will use RPA to derive and analyze equations for the
moments $M^{(p)}_k$ for arbitrary orders $p$ and thereby describe the
statistical evolution of the spectral fluctuations.  Note that RPA,
without a stronger Gaussianity assumption, is totally sufficient for
the WT closure at any order.  This allows us to study wavefields with
moments $M^{(p)}_k$ very far from their Gaussian values, which may
happen, for example, because of the choice of initial conditions or a
non-Gaussianity of the energy source in the system.

In \cite{Ben} non-Gaussian fields of a rather different kind were
considered.  Namely, statistically homogeneous wave fields were
considered in an infinite space which initially have decaying
correlations in the coordinate space and, therefore, smooth cumulants
in the k-space, e.g.
$$
\langle a_{k_1} a_{k_2} {\overline a_{k_3}} {\overline a_{k_4}}
\rangle = n_{k_1} n_{k_2} (\delta^{k_1}_{k_3} \delta^{k_2}_{k_4} + 
\delta^{k_1}_{k_4}\delta^2_3) + C_{123} \delta^{k_1+k_2}_{k_3 + k_4}, $$
where $C_{123}$ is a smooth function of ${\bf k_1}, {\bf k_2}, {\bf k_3}$ and
$\delta$'s now mean Dirac deltas. On the other hand, by taking the
large box limit it is easy to see that our expression (\ref{wick})
corresponds to a {\em singular} cumulant $C_{123} = Q_{k_1}/{\cal V}
\, \delta^{k_1}_{k_2} \delta^{k_1}_{k_3}$. It tends to zero when the
box volume ${\cal V}$ tends to infinity and yet it gives a finite
contribution to the waveaction fluctuations in this limit.\footnote{
Thus, assuming a finite box is an important intermediate step when
introducing the relevant to the fluctuations objects like $Q_k$.}
This singular cumulant corresponds to a small component of the
wavefield which is long-correlated, - the case not covered by the
approach of \cite{Ben}. On the other hand, it would be
straightforward to go beyond our RPA by adding a cumulant part of
the initial fields which tends to a smooth function of ${\bf k_1}, {\bf k_2},
{\bf k_3}$ in the infinite box limit (like in \cite{Ben}). However, such
cumulants would give a box-size dependent contribution to the
waveaction fluctuations which vanishes in the infinite box limit
(e.g.  it would change $\xi_k^2$ by $C_{kkk}/{\cal V}$). Thus, in
large boxes the waveaction fluctuation for the fields with smooth
cumulants is fixed at the same value as the for Gaussian fields,
$\xi_k = n_k$, and introduction of the singular cumulant is
essential to remove this restriction on the level of fluctuations.
On the other hand, the smooth part of the cumulant has no bearing on
the closure ( as shown in~\cite{Ben}) and on the large-box
fluctuation and, therefore, will be omitted in this manuscript for
brevity and clarity of the analysis.

\section{  Time-scale separation analysis }
Consider weakly nonlinear dispersive waves in a periodic box.  
Here  we consider quadratic nonlinearity and the linear dispersion
relations $\omega_k$ which allow three-wave interactions. Example of
such systems include surface capillary waves~\cite{Zakfil} and
internal waves in the ocean~\cite{LT}.  In Fourier space, the general
form for the Hamiltonian systems with quadratic nonlinearity looks as
follows,\footnote{We will follow the RPA approach as presented by
Galeev and Sagdeev \cite{GS} but deal with a slightly more general
case where the wave field is not restricted by the condition
$\overline a(k) = a(-k)$. We will also use elements of the technique
and notations of \cite{Ben}. }
\BEA
{\cal H} &=& \sum_{n=1}^\infty \omega_n|c_n|^2 +\epsilon
\sum_{l,m,n=1}^\infty \left( 
V^l_{mn} \bar c_{l} c_m  c_n\delta^l_{m+n}+c.c.\right),\CR
i\dot c_l &=&\frac{\partial {\cal H}}{\partial \bar c_l}, \ \ 
c_l=a_l e^{-i \omega_l t}, \CR
i \, \dot a_l &=& \epsilon \sum_{m,n=1}^\infty \left( V^l_{mn} a_{m}
a_{n}e^{i\omega_{mn}^l t} \, \delta^l_{m+n} 
\right.\CR && \left. \hspace{3cm}
+ 2 \bar{V}^{m}_{ln} \bar a_{n}
a_{m} e^{-i\omega^m_{ln}t } \, \delta^m_{l+n}\right),
\label{Interaction} \EEA 
where $a_n=a(k_n)$ is the complex wave amplitude in the interaction representation, 
$k_n = 2 \pi n/L $, $L $ is the box side length,
$n=(n_1,n_2)$ for 2D, or $ n=(n_1,n_2, n_3)$ in 3D, (similar for $k_l
$ and $ k_m$), $
\omega^l_{mn}\equiv\omega_{k_l}-\omega_{k_m}-\omega_{k_m}$ and
$\omega_l=\omega_{k_l}$ is the wave linear dispersion relation.  Here,
$V^l_{mn} \sim 1$ is an interaction coefficient and $\epsilon$ is
introduced as a formal small nonlinearity parameter.

In order to filter out fast oscillations at the wave period, let us
seek for the solution at time $T$ such that $2 \pi / \omega \ll T \ll
1/\omega \epsilon^2$.  The second condition ensures that $T$ is a lot
less than the nonlinear evolution time.  Now let us use a perturbation
expansion in small $\epsilon$,
$$a_l(T)=a_l^{(0)}+\epsilon a_l^{(1)}+\epsilon^2 a_l^{(2)}.$$
Substituting this expansion in (\ref{Interaction}) we get in
the zeroth order
$ a_l^{(0)}(T)=a_l(0)\label{definitionofa} $,
i.e. the zeroth order term is time independent. This corresponds to
the fact that the interaction representation wave amplitudes are
constant in the linear approximation.  For simplicity, we will write
$a^{(0)}_l(0)= a_l$, understanding that a quantity is taken at $T=0$
if its time argument is not mentioned explicitly.  The first order is
given by
\BEA
a^{(1)}_l (T) = -i \sum_{m,n=1}^\infty \left(   V^l_{mn}
a_m a_n \Delta^l_{mn} \delta^l_{m+n}\right.\CR\left.\hskip 4cm
+
2 \bar{V}^m_{ln}a_m\bar{a}_n \bar\Delta^m_{ln}\delta^m_{l+n}
\right),
\label{FirstIterate}
\EEA
where 
$ \Delta^l_{mn}=\int_0^T e^{i\omega^l_{mn}t}d t =
({e^{i\omega^l_{mn}T}-1})/{i \omega^l_{mn}}. \label{NewellsDelta}
$
Here we have taken into account that $a^{(0)}_l(T)= a_l$ and
$a^{(1)}_k (0)=0$.  

To calculate the second iterate, write
\BEA
i\dot{ a}^{(2)}_l  = \sum_{m,n=1}^\infty \Big[  
V^l_{mn}\delta^l_{m+n} e^{i \omega^l_{mn} t}
\left(a_m^{(0)} a_n^{(1)}+ a_m^{(1)} a_n^{(0)}\right)
\CR \hspace{3cm}+
2\bar{V}^m_{ln}\delta^m_{l+n} e^{ -i \omega^m_{ln} t}
\left(a_m^{(1)} \bar{a}_n^{(0)}+ a_m^{(0)} \bar{a}_n^{(1)}\right)
\Big].\CR
\label{SecondIterateTimeDerivative}
\EEA 
We now have to substitute (\ref{FirstIterate}) into
(\ref{SecondIterateTimeDerivative}) and integrate over time to obtain
\end{multicols}
\leftline{-------------------------------------------------------------------------}
\BEA a_l^{(2)} (T)  &=& \sum_{m,n, \mu, \nu=1}^\infty \left[ 2 V^l_{mn} 
\left(
-V^m_{\mu \nu}a_n a_\mu a_\nu E[\omega^l_{n \mu \nu},\omega^l_{mn}]
\delta^m_{\mu + \nu} 
-2 
\bar V^\mu_{m \nu}a_n a_\mu \bar a_\nu \bar 
E[\omega^{l \nu}_{n \mu},\omega^l_{mn}]\delta^\mu_{m + \nu}\right)
\delta^l_{m+n} \right.\CR && \left.
+ 2 
\bar V^m_{ln} 
 \left(
-V^m_{\mu \nu}\bar a_n a_\mu a_\nu E[\omega^{ln}_{\mu \nu},-\omega^m_{ln}]
\delta^m_{\mu + \nu} 
-
2 \bar V^\mu_{m \nu}\bar a_n a_\mu \bar a_\nu 
E[-\omega^\mu_{n \nu l},-\omega^m_{l n}]  \delta^\mu_{m + \nu}
\right) \delta^m_{l+ n} \right. \CR && \left.
+ 2 
\bar V^m_{ln} 
 \left(
\bar V^n_{\mu \nu}a_m \bar a_\mu  \bar a_\nu \delta^n_{\mu + \nu}
E[-\omega^m_{l\nu\mu},-\omega^m_{ln}] 
+
2  V^\mu_{n \nu}a_m \bar a_\mu  a_\nu E[\omega^{\mu l}_{\nu m},
-\omega^m_{ln}]\delta^\mu_{n + \nu}\right)\delta^m_{l+n}
\right],\CR\label{SecondIterate}
\EEA
\rightline{---------------------------------------------------------------------
----}
\begin{multicols}{2}
\noindent
where we used $a^{(2)}_k (0)=0$ and introduced 
$E(x,y)=\int_0^T \Delta(x-y)e^{i y t} d t .$

\section{ Statistical description  } 

Let us now develop a statistical
description applying RPA to the fields $a^{(0)}_k$. Since phases and
the amplitudes are statistically independent in RPA, we will first
perform average over the random phases (denoted as $\langle
... \rangle_{\psi}$) and then we average over amplitudes (denoted as
$\langle ... \rangle_{A}$) to calculate the moments,
$$ M^{(p)}_k(T)\equiv \langle |a_k(T)|^{2p}\rangle_{\psi,A}.\ \ \ 
p=1,\ 2, \, 3\ ...,  $$
First, let us calculate $|a_l(T)|^{2 p}$ as
\end{multicols}
\leftline{-------------------------------------------------------------------------}
\BEA|a_l(T)|^{2 p}
= \left(a_l^{(0)}+\epsilon a_l^{(1)}+\epsilon^2 a_l^{(2)}\right)^p
\left(\bar a_l^{(0)}+\epsilon \bar a_l^{(1)}+\epsilon^2
\bar a_l^{(2)}\right)^p= 
|a_l^{(0)}|^{2p} +
\epsilon p |a^{(0)}_l|^{2p-2}\left(a_l^{(0)}\bar a_l^{(1)}+  \bar a_l^{(0)}
a_l^{(1)} \right) +\CR
\epsilon^2|a_l|^{2p-4}\Big[ C^2_p(a_l^{(0)} \bar a_l^{(1)})^2 +
C^2_p(\bar a_l^{(0)} a_l^{(1)})^2 + p^2 |a_l^{(0)}|^2 |a_l^{(1)}|^2 +
p | a_l^{(0)} |^2 \left( a_l^{(0)} \bar a_l^{(2)}+
\bar a_l^{(0)} a_l^{(2)}\right)\Big] + ... ,\nonumber \EEA 
\rightline{---------------------------------------------------------------------
----}
\begin{multicols}{2}
\noindent
where $C^2_p$ is the binomial coefficient.

Up to the second power in $\epsilon$ terms, we have
\BEA \langle |a_l(T)|^{2p}\rangle_\psi= 
 |a_l|^{2p} + \CR
\epsilon^2 |a_l|^{2p-2} \left( p^2 \langle |a_l^{(1)}|^2 \rangle_\psi + p \, \langle a_l^{(0)} \bar
a_l^{(2)}+ \bar a_l^{(0)} a_l^{(2)}\rangle_\psi \right)\CR
\label{pmoment}\EEA
Here, the terms
proportional to $\epsilon$ dropped out after the phase averaging.
Further, we assume that there is no coupling to the $k=0$ mode, i.e.
$V^{k=0}_{k_1 k_2} = V^{k1}_{k_1 k=0}=0$, so that there is no 
contribution of the
term like $ \langle (a_l^{(0)} \bar a_l^{(1)})^2 \rangle_\psi $.
We now use (\ref{FirstIterate}) and (\ref{SecondIterate})
and the  averaging over the phases to obtain
\BEA
 \langle |a^{(1)}_l|^2\rangle_\psi = \nonumber\\
2 \sum_{m,n}^\infty \big[
|V^l_{mn}|^2 \delta^l_{m+n} |\Delta^l_{mn}|^2 
|a_m|^2 |a_n|^2  \nonumber \\ 
+2 |V^n_{lm}|^2 |\delta^n_{l+m}|^2 \Delta^n_{lm} 
 |a_n|^2  |a_m|^2 
\big],  \CR 
 \langle a^{(0)}_l \bar a^{(2)}_l+\bar a^{(0)}_l a^{(2)}_l
\rangle_\psi =
\nonumber\\
-8{ |a_l|^2}  \sum_{m,n}^\infty \big[
|V^l_{mn}|^2 \delta^l_{m+n} E(0,\omega^l_{mn})  |a_m|^2
\CR
+|V^n_{lm}|^2 \delta^n_{l+m} E(0,\omega^n_{lm}) (|a_m|^2- |a_n|^2) 
\big].  \nonumber \EEA
Let us substitute these expressions into (\ref{pmoment}), perform
amplitude averaging, take the large box limit\footnote{The large box
limit implies that sums will be replaced with integrals, the Kronecker
deltas will be replaced with Dirac's deltas,
$\delta^l_{m+n}\to\delta^l_{mn}/{\cal V}$, where we introduced
short-hand notation, $\delta^l_{mn}=\delta(k_l-k_m-k_n)$. Further we
redefine $M^{(p)}_k/{\cal V}^p \to M^{(p)}_k$.} and then large $T$
limit ($T \gg 1/ \omega$)\footnote{Note that
$\lim\limits_{T\to\infty}E(0,x)= T (\pi \delta(x)+iP(\frac{1}{x}))$,
and $\lim\limits_{T\to\infty}|\Delta(x)|^2=2\pi T\delta(x)$ (see e.g.
\cite{Ben}).}. We have
\BE
 M^{(p)}_k(T) =  M^{(p)}_k(0) + T
\left(-p \gamma_k M^{(p)}_k + p^2\eta_k M^{(p-1)}_k\right),
\EE
with
\BEA \eta_k = 4 \pi \epsilon^2 \int d {\bf k_1} d {\bf k_2} \, n_{1} n_{2} 
( 
|V^k_{12}|^2 \delta^k_{12} \delta(\omega^k_{12})
 \nonumber\\  \left.
+2 |V^2_{k1}|^2 \delta^2_{k1} \delta(\omega^2_{k1})
\right),  \label{RHO} \\
\gamma_k = 
8 \pi \epsilon^2 \int  d {\bf k_1} d {\bf k_2}   
(
|V^k_{12}|^2 \delta^k_{12} \delta(\omega^k_{12}) n_{2} 
\CR
 +|V^2_{k1}|^2 \delta^2_{k1} \delta(\omega^2_{k1}) (n_{1}- n_{2}) 
).  \label{GAMMA}
\EEA
Now, assuming that $T$ is a lot less than the nonlinear time ($T \ll
1/\omega \epsilon^2$) we finally arrive at our main result, 
\BE 
\dot M^{(p)}_k = -p \gamma_k M^{(p)}_k + 
p^2 \eta_k M^{(p-1)}_k.\label{MainResultOne}\EE
In particular, for the waveaction spectrum $M^{(1)}_k=n_k $
(\ref{MainResultOne}) gives the familiar kinetic equation (KE)
\BE \dot n_{k} = -\gamma_k n_{k} +\eta_k= \epsilon^2 J(n_{k}),
\label{KE1}\EE
 where $ J(n_{k})$ is the ``collision'' term \cite{ZLF,GS}, 
$$ J(n_{k})=\int d {\bf k_2} d {\bf k_1} (R^k_{12}-R^1_{k2}-R^2_{1k}), $$
 with 
\BE R_{k12}=4\pi|V^k_{12}|^2 \delta^k_{12}\delta(\omega^k_{12})
\Big(n_{2}n_{1}-n_{k}(n_{2}+n_{1})\Big). \EE
The second equation in the series (\ref{MainResultOne}) allows to
obtain the r.m.s. $\xi_k^2 = M^{(2)}_k - n^2_k$ (\ref{NazarenkoXI}) of
the fluctuations of the waveaction $\langle |a_k|^2\rangle$.  We
emphasize that (\ref{MainResultOne}) is valid even for strongly
intermittent fields with big fluctuations.

\section{Analysis of solutions: Gaussianity vs Intermittency}
Let us now consider the stationary solution of (\ref{MainResultOne}),
$ \dot M^{(p)}_k =0$ for all $p$. Then for $p=1$ from (\ref{KE1}) we
have $\eta_k=\gamma n_{k}$. Substituting this into
(\ref{MainResultOne}) we have
$$ M^{(p)}_k = p M^{(p-1)}_k n_{k}\label{ResultTwo}, $$
with the solution
$ M_k^{(p)}=p!\ n_{k}^p$.
Such a set of moments correspond to a Gaussian wavefield $a_k$.  To
see how such a Gaussian steady state forms in time, let us rewrite
(\ref{MainResultOne}) in terms of relative deviations of $M^{(p)}_k$
from their Gaussian values,
%
%
$$F^{(p)}_k = {M^{(p)}_k - p! \ n^p_k \over p! \ n^p_k}, \,\,\,
 p=1,2,3,... .$$
By definition, $F_k^{(1)}$ is always zero. For $p=2$, this expression
measures the flatness of the distribution of Fourier amplitudes at
each $k$.  This quantity determines the r.m.s. $\xi_k^2 = M^{(2)}_k -
n^2_k$ of the fluctuations of the waveaction $\langle |a_k|^2\rangle$,
(\ref{NazarenkoXI}) or the mean level of ``noisyness'',
$$\xi_k^2 =  n_k^2 (2 F^{(2)}_k +1). $$
%
Using (\ref{MainResultOne}), we  obtain 
\BE \dot F^{(p)}_k = 
{ p \eta_k \over n_k} ( F^{(p-1)}_k - F ^{(p)}_k),
\label{MainResultTwo}
\EE 
for $p=2,3,4,...$ This results has a particularly simple form of a
decoupled equation for $p=2$,
$$
\dot F^{(2)}_k = - { 2 \eta_k \over n_k} F^{(2)}_k.
$$ 
Taking into account that $\eta_k >0$, we see from this equation that
deviations of the {\em mean} level of fluctuations from Gaussianity
always decay. In fact, deviations $F^{(p)}$ decay at each {\em fixed}
$p$. This is easy to see from the general solution of
(\ref{MainResultTwo}) (obtained recursively):
\BE F^{(p)}_k(t) = e^{-p\theta} \sum^p_{j=2} {\theta^{p-j} p!  \over j!
(p-j)!}  F_k^{(j)}(t=0),
\label{fsoln}
\EE
where $\theta = \int_0^t {\eta_k \over n_k} dt'$ is a ``renormalized''
time variable. One can see that this expression decays exponentially
as $t \to \infty$ for any fixed $p$.

However, an interesting picture emerges at high $p$ corresponding to
high wave amplitudes.  Although the deviations $F^{(p)}_k$ eventually
decay at each {\em fixed} $p$, their initial values propagate in $p$
without decay toward the larger values of $p$.  Indeed, one can
approximate (\ref{MainResultTwo}) for $p \gg 1$ by a first-order PDE,
$$
\partial_t F^{(p)}_k + {p \eta_k \over n_k} \partial_p F^{(p)}_k =0.
$$
According to this equation, $F^{(p)}_k$ propagates toward high $p$'s
as a wave. This wave does not change shape with respect to coordinate
$x= \ln p$ and, therefore, it spreads in $p$ without change in
amplitude. The speed of this wave (in $x$) is time independent for
statistically steady states (i.e. when $n_k$ and $\eta_k$ are time
independent). Note that this dynamics occurs at each $k$ practically
independently, i.e. the only coupling of different $k$'s occurs in the
propagation speed via $\eta_k$.

These solutions allow us to establish the character of intermittency
in wave turbulence systems, i.e. to describe how high-amplitude
``bursts'' occur with greater than Gaussian probabilities.  In terms
of the PDF, the wave of non-Gaussianity $F^{(p)}_k$ toward high values
of $p$ corresponds to a wave propagating from low-amplitude ``bulk''
part to the high-amplitude ``tail'' on the PDF profile. Indeed, a
Gaussian PDF for $a_k$ corresponds to a distribution of $\lambda=
|a_k|^2$ of form $P(\lambda) = n^{-1} e^{-\lambda/n}$, and moment
$M^{(p)}_k$ ``probes'' this distribution in a range of $\lambda$
around $\lambda_p = p n$ with a characteristic width $\delta \lambda
\sim n$. Lifting $P(\lambda)$ in this range by a certain factor will
result in an increase of moment $M^{(p)}_k$ by the same factor. Thus,
the wave propagating from small to large $p$'s corresponds to a wave
from low to high $\lambda$'s.  This wave is such that the relative
deviation from distribution $P(\lambda) =n^{-1} e^{-\lambda/n}$
remains unchanged, but the range of $\lambda$'s at which such
non-Gaussianity occurs moves into the tail (with speed $\eta$) and
spreads (proportionally to its position in $\lambda$).  Note however
that at each {\em fixed} $\lambda$ deviations from $P(\lambda) =
n^{-1} e^{-\lambda/n}$ decay, which corresponds to decay of $F^{(p)}$
at each fixed $p$ at large time.

Predictions (\ref{fsoln}) about the behavior of fluctuations of the
waveaction spectra can be tested by modern experimental techniques
which allow to produce surface water waves with random phases and a
prescribed shape of the amplitude $|a_k|$ \cite{lev}.  It is even
easier to test (\ref{fsoln}) numerically.  Consider for example
capillary waves on deep water.  If a Gaussian forcing at low $k$
values is present, the steady state solution of the kinetic equation
corresponds to the Zakharov-Filonenko (ZF) spectrum of Kolmogorov
type~\cite{ZLF,Zakfil}. It is given by
\BE n_k = A k^{-17/4} \label{ZF},\EE
with $A=\sqrt{P} \rho^{3/2} C/\sigma^{1/4}$, where $P$ is the value of
flux of energy toward high wavenumbers, $\rho$ and $\sigma$ are the
density and surface tension of water, and $C\simeq 13.98$. The
simplest experiment would be to start with a zero-fluctuation
(deterministic) spectrum and to compare the fluctuation growth with
the predictions of (\ref{MainResultOne}).  Note that such
no-fluctuations initial conditions were used in
\cite{OsbornePRL,ZakharovPRL}.

Let us calculate the rate at which a fluctuations grow for such an
initial conditions. Since $n_k$ and $\eta_k$ are time independent in
this case, we have $\theta = \eta_k t/n_k =\gamma_k t$.  Thus, the
only quantity we need to calculate is $\gamma_k$.  Let us take into
account that the spectrum $n_k$ is isotropic, that is it depends only
on the modulus of the vector, not on its directions. We then can
perform an angular averaging of (\ref{GAMMA}) obtaining
\BEA
\gamma_k = 
8 \epsilon^2 \int  d {\bf k_1} d {\bf k_2} S_{k k_1 k_2}^{-1}  
(
|V^k_{12}|^2  \delta(\omega^k_{12}) n_{2} 
\CR
 +|V^2_{k1}|^2  \delta(\omega^2_{k1}) (n_{1}- n_{2}) 
),
\nonumber\\ 
S_{k 1 2} = \left< \delta({\bf k}-{\bf
k_1}-{\bf k_2})\right>\equiv {1 \over 4 \pi^2} \int \delta({\bf
k}-{\bf k_1}-{\bf k_2}) \, d \theta_1 d \theta_2  \nonumber\\
= \frac{1}{2}\sqrt{
2 \left( (k k_1)^2 +(k k_2)^2 +(k_1 k_2)^2 
\right)-k^4-k_1^4 -k_2^4} \, . \nonumber\\
 \label{GAMMAdeltaAVERAGE}
\EEA

Let us substitute ZF spectrum (\ref{ZF}) into
(\ref{GAMMAdeltaAVERAGE}), take the values of $\omega_k$ and
$V^k_{12}$ appropriate for the capillary waves on deep
water(\cite{ZLF}, eqs (5.2.1-2)). By changing the variables of
integrations via $k_1= k \xi_1, \ \ k_2 = k \xi_2$ we can factor out
the $k$ dependence of $\gamma_k$. Performing one of $\xi$ integrals
analytically with the use of the delta function in $\omega$'s, we
perform the remaining single integral numerically to obtain (all the
integrals converge):
$${\gamma= \frac{4.30 A \sqrt{\sigma}}{16 \pi \rho^{3/2}} k^{3/4}},
$$
where the dimensionless constant $4.30$ was obtained by numerical
integration. Substituting the value of $A$ we finally obtain
$${\gamma= {1.20 \sqrt{P} \sigma^{1/4} } k^{3/4}},
$$
Consequently, our prediction for the fluctuations growth is
\BEA
\xi_k^2 = n_k^2 (2 F^{(2)}_k +1) = A^2 k^{-17/2} (1- e^{-2 \gamma_k t}).
\EEA
%

%
%
Note that fluctuations stabilize at Gaussian values faster for high
$k$ values.  One can also substitute $\gamma_k$ calculated for the
capillary waves into the solutions for the higher $p$'s,
(\ref{fsoln}). Again, the dynamics here is going to be faster at large
$k$'s because they correspond to higher values of $\theta =\gamma t$.
In particular, at large $k$ there will be a faster wave toward higher
$p$'s. For the particular type of initial conditions we have taken (no
initial fluctuations), this wave will describe formation of a Rayleigh
distribution (corresponding to the Gaussian statistics of $a_k$)
behind a propagating front on the PDF profile.  It a way, this
dynamics is non-intermittent: zero initial fluctuations grow to the
Gaussian level but never exceed it.

It is also interesting to test our predictions when the initial
conditions, or forcing, are non-Gaussian, as in most practical
situations.  Our theory predicts that non-Gaussianity of the
low-amplitude (bulk) part of the PDF will propagate without decay into
the high-amplitude tail at each fixed $k$. The speed of this
propagation is proportional to $\gamma_k$ and, therefore, will be
higher for large $k$'s in the case of the capillary waves.  This means
higher intermittency in the low-$k$ range in the case of stationary
forced turbulence.

\section{ Discussion } 

In this manuscript, we derived a hierarchy of
equations~(\ref{MainResultOne}) for the one-point moments $M^{(p)}_k$
of the waveaction $|a_k|^2$. This system of equations has a
``triangular'' structure: the time derivative of the $p$-th moment
depends only on the moments of order $p, p-1$ and 1 (spectrum). Their
evolution is not ``slaved'' to the spectrum or any other low moments
and it depends on the initial conditions. RPA allows the initial
conditions to be far from Gaussian and deviation of n'th moment from
its Gaussian.  Among two allowed extreme limits are the wavefield with
a deterministic amplitude $|a_k|$ (for which $M^{(p)}_k = n^p_k$) and
the intermittent wavefields characterized by sparse k-space
distributions of $|a_k|$ (for which $M^{(p)}_k \gg n^p_k$).

Equations~(\ref{MainResultTwo}) for the deviations from Gaussianity
have an interesting property that the nonlinear coupling between
different modes $k$ occurs only via a rate constant $\eta/n$. By
removing this dependence into a ``renormalized'' time $\theta$ one
gets a linear system of equation which can be easily solved in the
general case, see (\ref{fsoln}).  Analyzing these solutions we showed
that the deviation from Gaussianity decreases as at each fixed
amplitude $|a_k|$. At the same time, we showed that any initial
non-Gaussianity at small amplitudes propagates as a non-decaying wave
toward the high-amplitude tail of the PDF.  This process describes the
character of the wave turbulence intermittency when high-amplitude
wave ``bursts'' occur in the system more frequently than predicted for
Gaussian fields.  On the other hand, the assumption about the weak
nonlinearity breaks down when a high amplitude burst occurs in the
system, leading to a failure of the RPA closure to describe the PDF
tails.  One can conjecture that the resulting phase coherence will
lead to a nonlinear amplitude saturation which will stop the wave
predicted by our theory which, in turn, will lead to a stagnation and
accumulation in this region on the PDF tail.  Thus, it is natural to
expect even stronger intermittency when the higher order nonlinear
effects are taken into account.

We would like to emphasize that the type of intermittency discussed in
the present manuscript appears within the weakly nonlinear closure and
not as a result of its breakdown as in \cite{NazarenkoNewell}. This
intermittency is quite subtle and it occurs only in the PDF tails and
not in its core (which tends to a Gaussian state).  As a result, the
lower moments will not feel these rare ``bursts'' and they will evolve
as predicted by the WT closure. We have showed that this kind of
intermittency inevitably occurs at {\em all} wavenumbers $k$ provided
some initial non-Gaussianity is present in the PDF core.  Paper
\cite{NazarenkoNewell} considers a different and a more dramatic kind
of intermittency which occurs simultaneously with the strong
nonlinearity of the typical wave from the PDF core. This kind of
intermittency is more seldom and it takes place only in some special
parts of the $k$-space (e.g. at very small scales).  In particular, it
never occurs for the capillary waves considered in this paper provided
that only weakly nonlinear waves are produced at the forcing scale.

The present paper deals with the three-wave systems only.  The
four-wave resonant interactions are slightly more complicated in that
the nonlinear frequency shift occurs at a lower order in nonlinearity
parameter than the nonlinear evolution of the wave amplitudes. To
build a consistent description of the amplitude moments one has to
perform a renormalization of the perturbation series taking into
account the nonlinear frequency shift.  This will be done in a future
publication.

{\bf Acknowledgments} Authors thank Alan Newell for enlightening
discussions. YL is supported by NSF CAREER grant DMS 0134955 and by
ONR YIP grant N000140210528. SN thanks ONR for the support of his
visits to RPI.

\end{multicols}
\end{document}